\documentclass[a4paper]{PoS}
\usepackage[utf8]{inputenc}
\usepackage{url}
\usepackage{amssymb}          
\usepackage[amssymb]{SIunits}
\usepackage{amsmath} 
\usepackage{graphicx}
\usepackage{subfigure}
\usepackage{color}

\usepackage{tabularx}
\usepackage{booktabs}
\usepackage{longtable}
\usepackage{multirow}

\usepackage{tikz}
\usetikzlibrary{calc,shapes,arrows}
\usetikzlibrary{decorations.pathmorphing}
\usetikzlibrary{decorations.markings}
\usetikzlibrary{positioning}

\newcommand{\elepton}{\ensuremath{\textrm{e}}}
\newcommand{\electron}{\ensuremath{\elepton^-}}
\newcommand{\positron}{\ensuremath{\elepton^+}}

\newcommand{\proton}{\ensuremath{\textrm{p}}}
\newcommand{\neutron}{\ensuremath{\textrm{n}}}

\title{High energy astroparticle physics for high school students}

\ShortTitle{Astroparticle Masterclass}
  
\author{\speaker{Maria Krause}$^{a}$, Hans-Peter Bretz$^{a}$, Lew Classen$^{b}$, Markus Holler$^{c}$, Moritz H\"{u}tten$^{a}$, Susanne Raab$^{b}$, Julian Rautenberg$^{d}$, Anneli Schulz$^{a}$ \\
\llap{$^{a}$}DESY \\ Platanenallee 6, 15738 Zeuthen, Germany \\
\llap{$^{b}$}Universit\"{a}t Erlangen-N\"{u}rnberg, Physikalisches Institut \\ Erwin-Rommel-Str. 1, 91058 Erlangen, Germany \\ 
\llap{$^{c}$}Laboratoire Leprince-Ringuet, Ecole Polytechnique, CNRS/IN2P3 \\ 91128 Palaiseau, France \\ 
\llap{$^{d}$}Bergische Universit\"{a}t Wuppertal\\ Gau\ss stra\ss e 20, 42119 Wuppertal, Germany\\
E-mail: \email{maria.krause@desy.de}}

\abstract{The questions about the origin and type of cosmic particles are not only fascinating for scientists in astrophysics, but also for young enthusiastic high school students. To familiarize them with research in astroparticle physics, the Pierre Auger Collaboration agreed to make 1\% of its data publicly available. The Pierre Auger Observatory investigates cosmic rays at the highest energies and consists of more than 1600 water Cherenkov detectors, located near Malarg\"{u}e, Argentina. With publicly available data from the experiment, students can perform their own hands-on analysis. In the framework of a so-called \textit{Astroparticle Masterclass} organized alongside the context of the German outreach network \textit{Netzwerk Teilchenwelt}, students get a valuable insight into cosmic ray physics and scientific research concepts. We present the project and experiences with students.}

\FullConference{The 34th International Cosmic Ray Conference,\\
		30 July- 6 August, 2015\\
		The Hague, The Netherlands}

\begin{document}

\section{Introduction}
The project \textit{Netzwerk Teilchenwelt} \cite{NTWurl} is a network of communication specialists, science edu\-cators, scientists, and researchers. It consists of 24 German research institutes in particle and astroparticle physics with the aim of enabling students to authentically experience modern physics research and to interact with more than 100 researchers. High school students from 15 to 19 years old have the opportunity to be a scientist for one day during which they perform their own measurements and analyses of astroparticle physics data.

Within the framework of this project the \textit{Astroparticle Masterclasses} are developed as an edu\-cational project with the primary goal of bringing experimental data and research methods into the classroom. High school students get the chance to explore the fascinating world of astroparticle physics, combined with the experience of how scientists investigate nature. Researchers and scientists are invited to spend a day at a participating school. There, students and teachers get acquainted with methods and questions of current scientific research. The highlight of this day is the performance of measurements on real data from astroparticle physics experiments. The project is based on the public dataset from the Pierre Auger Observatory \cite{Aab:2015a, PAOpublicDE, PAOpublicINT}.

The following sections will cover a short and basic introduction to the physics of extensive air showers (\S \ref{sec:EAS}), which are detected at the Pierre Auger Observatory (\S  \ref{sec:PAO}). This will be followed by discussing feedback with participants on improvements that can be made for future events (\S \ref{sec:activity} and \ref{sec:discussion}).

\section{Physics of Extensive Air Showers}
\label{sec:EAS}
If a high-energy particle such as a proton or iron nucleus hits the top of the atmosphere, it interacts with the nucleus of an atom in the air. This primary interaction produces a jet of secondary particles which then results in successive interactions due to a chain reaction of particles. This phenomenon is called an extensive air shower (EAS) \cite{GriederI}, and is schematically illustrated in \mbox{Figure \ref{fig: EAS}}. The development of each air shower stops when the energy of the produced particle is too small to produce secondary particles. A shower front is formed by millions of secondary particles, which spread out at nearly the speed of light, and is a few meters thick at the center. A fraction of the secondary particles in the EAS propagates down to the ground and penetrates the surface detectors of the Pierre Auger Observatory. Due to the large amount of particles in the EAS, the sample of particles measured on the ground is a good representation of the air shower. 

\tikzset{%
  primary/.style={postaction={decorate}, decoration={markings,mark=at position .5 with {\arrow{>}}}},%
  hadron/.style={},
  meson/.style={},
  lepton/.style={},
  photon/.style={decorate, decoration={snake, amplitude=.4mm, segment length=1.5mm}},
  neutrino/.style={dashed},
  synch/.style={color=gray!40,fill=gray!40}
}
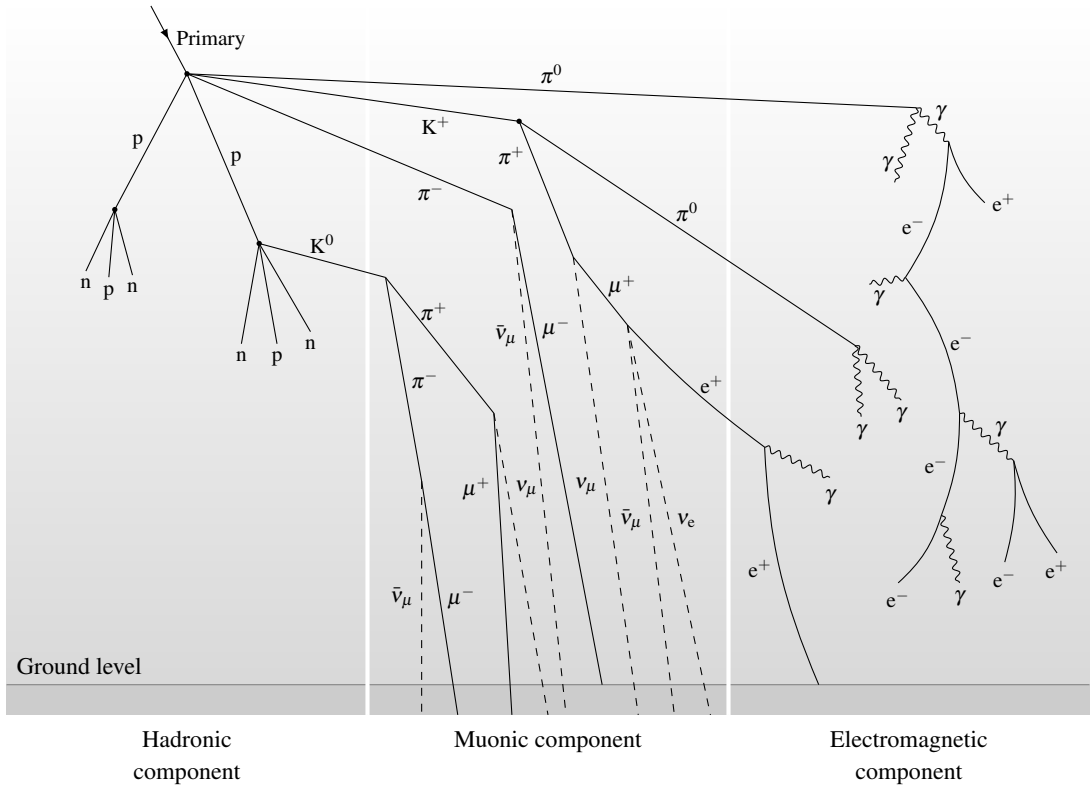
\begin{figure}[t]
\centering
\begin{tikzpicture}[>=latex]
  \def\asch{9} 
  \def\ascw{4.75} 
  \coordinate (Left) at (0,0);
  \coordinate (Right) at ($({3*\ascw},0)$);
  \coordinate (Top) at (0,\asch);
  \coordinate (Surface) at (0,0);
  \coordinate (Bottom) at (0,-.4);

  \fill[black!20] ($(Left) + (Surface)$) rectangle ($(Right) + (Bottom)$);

  \shade[bottom color=gray!30, top color=gray!0] ($(Top)+(Left)$) rectangle ($(Surface)+(Right)$);

  \draw[color=black!50] ($(Left) + (Surface)$) node[above right, text=black]{\footnotesize Ground level} -- ($(Right) + (Surface)$);

  \coordinate (ASC0) at ($(0*\ascw,0)$);
  \def\ascb{0.02}
  \foreach \i in {1,2} {
    \coordinate (ASC\i) at ($(\i*\ascw,0)$);
    \draw[color=white, fill=white] ($(ASC\i)+(Top)+(-\ascb,0)$) rectangle ($(ASC\i)+(Bottom)+(\ascb,0)$);
  }

  \newcommand\SyncRad[4]{%
    \coordinate (#1SL) at ($ (#1) + ({#3 + #4}:#2) $);
    \coordinate (#1SR) at ($ (#1) + ({#3 - #4}:#2) $);
    \draw[synch] (#1)--(#1SL)--(#1SR)--cycle;
  }

  \footnotesize
  \node[below, text width=2.5cm, text badly centered] (HComp) at ($(Left)! 0.5 !(ASC1) ! .5cm ! -90:(ASC1) $) {Hadronic component};
  \node[below, text width=2.5cm, text badly centered] (MComp) at ($(ASC1)! 0.5 !(ASC2) ! .5cm ! -90:(ASC2) $) {Muonic component};
  \node[below, text width=2.5cm, text badly centered] (EComp) at ($(ASC2)! 0.5 !(Right)! .5cm ! -90:(Right)$) {Electromagnetic component};

  \scriptsize
  \coordinate (Primary) at ($(ASC0) + (Top) + (0.4*\ascw,0)$);
  \coordinate (H0) at ($(Primary) + (.1*\ascw,-0.1*\asch)$);
  \coordinate (H01) at ($(H0) + (-0.2*\ascw,-0.2*\asch)$);
  \coordinate (H02) at ($(H0) + (0.2*\ascw,-0.25*\asch)$);
  \coordinate (M0) at ($(ASC1) + (Top) + (0.42*\ascw,-0.17*\asch)$);
  \coordinate (M1) at ($(ASC1) + (Top) + (0.4*\ascw,-0.3*\asch)$);
  \coordinate (M2) at ($(ASC1) + (Top) + (0.05*\ascw,-0.4*\asch)$);
  \coordinate (M00) at ($(M0) + (0.15*\ascw,-0.2*\asch)$); 
  \coordinate (M000) at ($(M00) + (0.15*\ascw,-0.1*\asch)$); 
  \coordinate (M20) at ($(M2) + (0.3*\ascw,-0.2*\asch)$); 
  \coordinate (M21) at ($(M2) + (0.1*\ascw,-0.3*\asch)$); 
  \coordinate (E0) at ($(ASC2) + (Top) + (0.52*\ascw,-0.15*\asch)$);
  \coordinate (E00) at ($(E0) + (0.09*\ascw, -0.05*\asch)$);
  \coordinate (E01) at ($(E0) + (-0.06*\ascw, -0.11*\asch)$);
  \coordinate (E000) at ($(E00) + (-0.12*\ascw, -0.2*\asch)$);
  \coordinate (E001) at ($(E00) + (0.1*\ascw, -0.09*\asch)$);
  \coordinate (E0000) at ($(E000) + (0.15*\ascw, -0.2*\asch)$);
  \coordinate (E0001) at ($(E000) + (-0.1*\ascw, -0.01*\asch)$);
  \coordinate (E00000) at ($(E0000) + (-0.05*\ascw, -0.15*\asch)$);
  \coordinate (E00001) at ($(E0000) + (0.15*\ascw, -0.07*\asch)$);
  \coordinate (E000000) at ($(E00000) + (-0.12*\ascw, -0.10*\asch)$);
  \coordinate (E000001) at ($(E00000) + (0.05*\ascw, -0.10*\asch)$);
  \coordinate (E000010) at ($(E00001) + (-95:0.15*\asch)$);
  \coordinate (E000011) at ($(E00001) + (-65:0.15*\asch)$);
  \coordinate (E1) at ($(ASC2) + (Top) + (0.2*\ascw, -0.23*\asch)$);
  \coordinate (E10) at ($(E1) + (-45:0.2*\ascw)$);
  \coordinate (E11) at ($(E1) + (-75:0.2*\ascw)$);
  \coordinate (E2) at ($(ASC2) + (Top) + (0.35*\ascw, -0.50*\asch)$);
  \coordinate (E20) at ($(E2) + (-50:0.2*\ascw)$);
  \coordinate (E21) at ($(E2) + (-85:0.2*\ascw)$);
  \coordinate (E3) at ($(ASC2) + (Top) + (0.1*\ascw, -0.65*\asch)$);
  \coordinate (E30) at ($(ASC2) + (0.25*\ascw,0) + (Surface)$);
  \coordinate (E31) at ($(E3) + (-25:0.2*\ascw)$);

  \foreach \point in {H0,H01,H02,M0} {
    \fill[black] (\point) circle (1pt);
  }

  \draw[primary] (Primary) -- node[right]{Primary} (H0);
  \draw[hadron] (H0) -- node[left] {p} (H01);
  {
    \draw[hadron] (H01) -- +(-115:0.1*\asch) node[below] {\neutron};
    \draw[hadron] (H01) -- +(-95:0.1*\asch) node[below] {\proton};
    \draw[hadron] (H01) -- +(-75:0.1*\asch) node[below] {\neutron};
  }
  \draw[hadron] (H0) -- node[right] {p} (H02);
  {
    \draw[hadron] (H02) -- +(-100:.15*\asch) node[below] {\neutron};
    \draw[hadron] (H02) -- +(-80:.15*\asch) node[below] {\proton};
    \draw[hadron] (H02) -- +(-60:.15*\asch) node[below] {\neutron};
    \draw[meson] (H02) -- node[above]{$\mathrm{K}^0$}(M2);
    {
      \draw[meson] (M2) -- node[above, right, near start]{$\pi^+$}(M20);
      {
        \draw[lepton] (M20) -- node[left, near start]{$\mu^+$}($(ASC1) + (0.4*\ascw,0) + (Bottom)$);
        \draw[neutrino] (M20) -- node[right, near start]{$\nu_\mu$}($(ASC1) + (0.5*\ascw,0) + (Bottom)$);
      }
      \draw[meson] (M2) -- node[right]{$\pi^-$}(M21);
      {
        \draw[neutrino] (M21) -- node[left]{$\bar\nu_\mu$}($(ASC1) + (0.15*\ascw,0) + (Bottom)$);
        \draw[lepton] (M21) -- node[right]{$\mu^-$}($(ASC1) + (0.25*\ascw,0) + (Bottom)$);
      }
    }
  }
  \draw[meson] (H0) -- node[below, near end]{$\pi^-$}(M1);
  {
    \draw[neutrino] (M1) --  node[left, near start]{$\bar\nu_\mu$}($(ASC1) + (0.55*\ascw,0) + (Bottom)$);
    \draw[lepton] (M1) -- node[right, near start]{$\mu^-$}($(ASC1) + (0.65*\ascw,0) + (Surface)$);
  }
  \draw[meson] (H0) -- node[below, near end]{$\mathrm{K}^+$}(M0);
  {
    \draw[meson] (M0) -- node[above=.5ex]{$\pi^0$} (E2);
    {
      \draw[photon] (E2) -- node[below, at end]{$\gamma$} (E20);
      \draw[photon] (E2) -- node[below, at end]{$\gamma$} (E21);
    }
    \draw[meson] (M0) -- node[left, near start]{$\pi^+$}(M00);
    {
      \draw[neutrino] (M00) -- node[left]{$\nu_\mu$}($(ASC1) + (0.75*\ascw,0) + (Bottom)$);
      \draw[lepton] (M00) -- node[above=.5ex,right]{$\mu^+$}(M000);
      {
        \draw[lepton] (M000) edge[bend right=5] node[above=.5ex, right]{\positron} (E3);
        {
          \draw[lepton] (E3) edge[bend right=10] node[left]{\positron} (E30);
          \draw[photon] (E3) -- node[below=.5ex, at end]{$\gamma$} (E31);
        }
        \draw[neutrino] (M000) -- node[left]{$\bar\nu_\mu$}($(ASC1) + (0.85*\ascw,0) + (Bottom)$);
        \draw[neutrino] (M000) -- node[right]{$\nu_\mathrm{e}$}($(ASC1) + (0.95*\ascw,0) + (Bottom)$);
      }
    }
  }
  \draw[meson] (H0) -- node[above]{$\pi^0$} (E0);
  {
    \draw[photon] (E0) -- node[above=.5ex, near end]{$\gamma$} (E00);
    {
      \draw[photon] (E0) -- node[left, near end]{$\gamma$} (E01);
    }
    \draw[lepton] (E00) edge[bend left=15] node[left, pos=0.6]{\electron}  (E000);
    {
      \draw[lepton] (E00) edge[bend right=15] node[right, pos=1.0]{\positron} (E001);
    }
    \draw[lepton] (E000) edge[bend left=15] node[right]{\electron} (E0000);
    {
      \draw[photon] (E000) -- node[below, near end]{$\gamma$} (E0001);
    }
    \draw[lepton] (E0000) edge[bend left=10] node[left]{\electron}(E00000);
    {
      \draw[lepton] (E00000) edge[bend left=15] node[below, at end]{\electron} (E000000);
      \draw[photon] (E00000) -- node[below, at end]{$\gamma$} (E000001);
    }
    \draw[photon] (E0000) -- node[above=.5ex, near end]{$\gamma$}(E00001);
    {
      \draw[lepton] (E00001) edge[bend left=10] node[below, at end]{\electron} (E000010);
      \draw[lepton] (E00001) edge[bend right=10] node[below, at end]{\positron} (E000011);
    }
  }

\end{tikzpicture}
\caption{Sketch of the longitudinal development of an extensive air shower induced by a primary particle. The shower includes the hadronic, muonic, and electromagnetic component as well as neutrinos (adapted from \cite{Lafebre}).}
\label{fig: EAS}
\end{figure}


\section{Pierre Auger Observatory}
\label{sec:PAO}
The Pierre Auger Observatory is located on the Pampa Amarilla outside the city of Malarg\"{u}e in Argentina at an elevation of about \unit{1400}{m} above sea level. This ultra-high energy cosmic-ray experiment addresses several unresolved questions about the spectrum, origin, composition, and interactions of cosmic particles possessing energies up to \unit{10^{20}}{eV}. Several important discoveries in this research field have been already made \cite[and references therein]{Aab:2015a, Aab:2015kma}.

\begin{figure}[t]
 \centering
 \subfigure{\includegraphics[height=6cm,keepaspectratio]{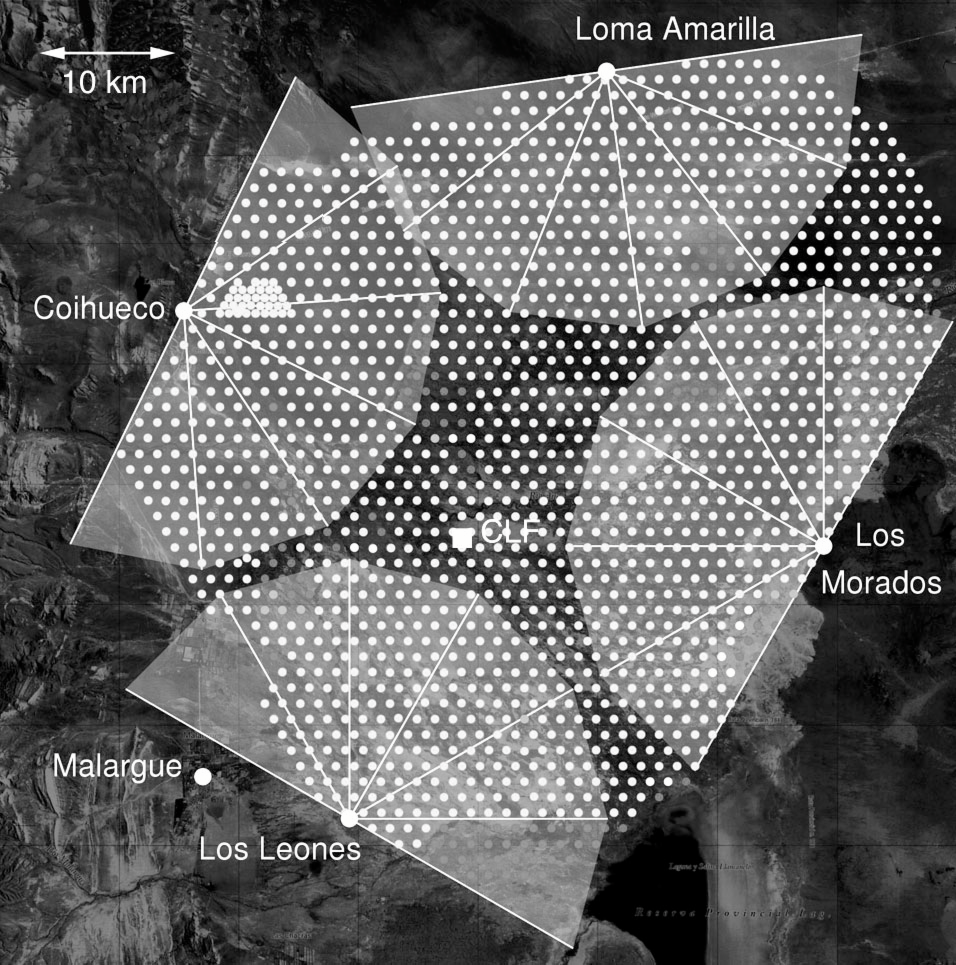}}
\quad
 \subfigure{\includegraphics[height=6cm,keepaspectratio]{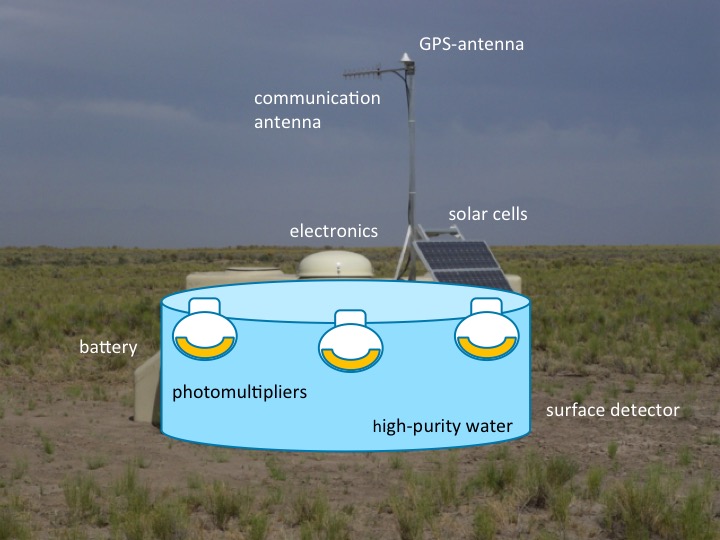}}
 \caption{\textbf{Left:} The Pierre Auger Observatory \cite{Abraham:2009pm}. Every white dot represents one surface detector. The white lines show the field of view of the fluorescence detectors. \textbf{Right:} Components of a surface detector station of the Pierre Auger Observatory (adapted from \cite{Aab:2015a}).}
 \label{fig:PAO}
\end{figure}

The layout of the observatory is shown in the left panel of Figure \ref{fig:PAO}. The observatory applies two detection techniques. For the purpose of the \textit{Astroparticle Masterclass} only data from the surface detector stations are used. There are 1660 water-Cherenkov particle detector stations which together comprise the surface detector covering an area of about \unit{3000}{km^{2}}.  A water-Cherenkov detector (right-hand plot of Figure \ref{fig:PAO}) is \unit{1.20}{m} high and optically opaque. It is covered by plastic material and filled with high-purity water. A charged particle emits Cherenkov photons when penetrating through the water if its velocity $v$ is higher than the speed of light in this medium: $\frac{v}{c}>\frac{1}{n}$, where $c$ is the velocity of light in vacuum and $n$ is the refractive index of the medium. Thus, the charged secondary particles of an extensive air shower, which travel with a velocity close to the speed of light, generate a light signal in water. Three photomultiplier tubes are mounted on the surface of the detectors. The Pierre Auger Collaboration uses a measurement unit named VEM (Vertical Equivalent Muon) to quantify the signal measured in each station. This means that the signal from a shower particle piercing the station is expressed in multiples of the intensity of the signal left by a single muon striking the station vertically. The measured signal is proportional to the total energy of the particles which entered the detector, which is thus proportional to the number of particles. The higher the energy of the primary particle, the more particles reach the detector. This information is essential to understand the masterclass activities. Using the measured particle density, the location of the shower axis on the ground can be evaluated.

\section{Description of the activity}
\label{sec:activity}
The purpose of the \textit{Astroparticle Masterclass} is that each student reconstructs an event measured by the Pierre Auger Observatory \cite{Abreu:2014a, deSouza:2015a}. The activity has been designed to be inexpensive and easy reproducible with little prior knowledge. The derivation of the properties of air showers will be done with a common spreadsheet program. The project is divided into three main parts. First, the impact point of the shower core will be reconstructed. Then, the arrival direction of the shower will be evaluated, and lastly, the lateral distribution of the air shower will be analyzed.

\begin{figure}[t]
 \centering
 \subfigure{\includegraphics[width=0.48\textwidth]{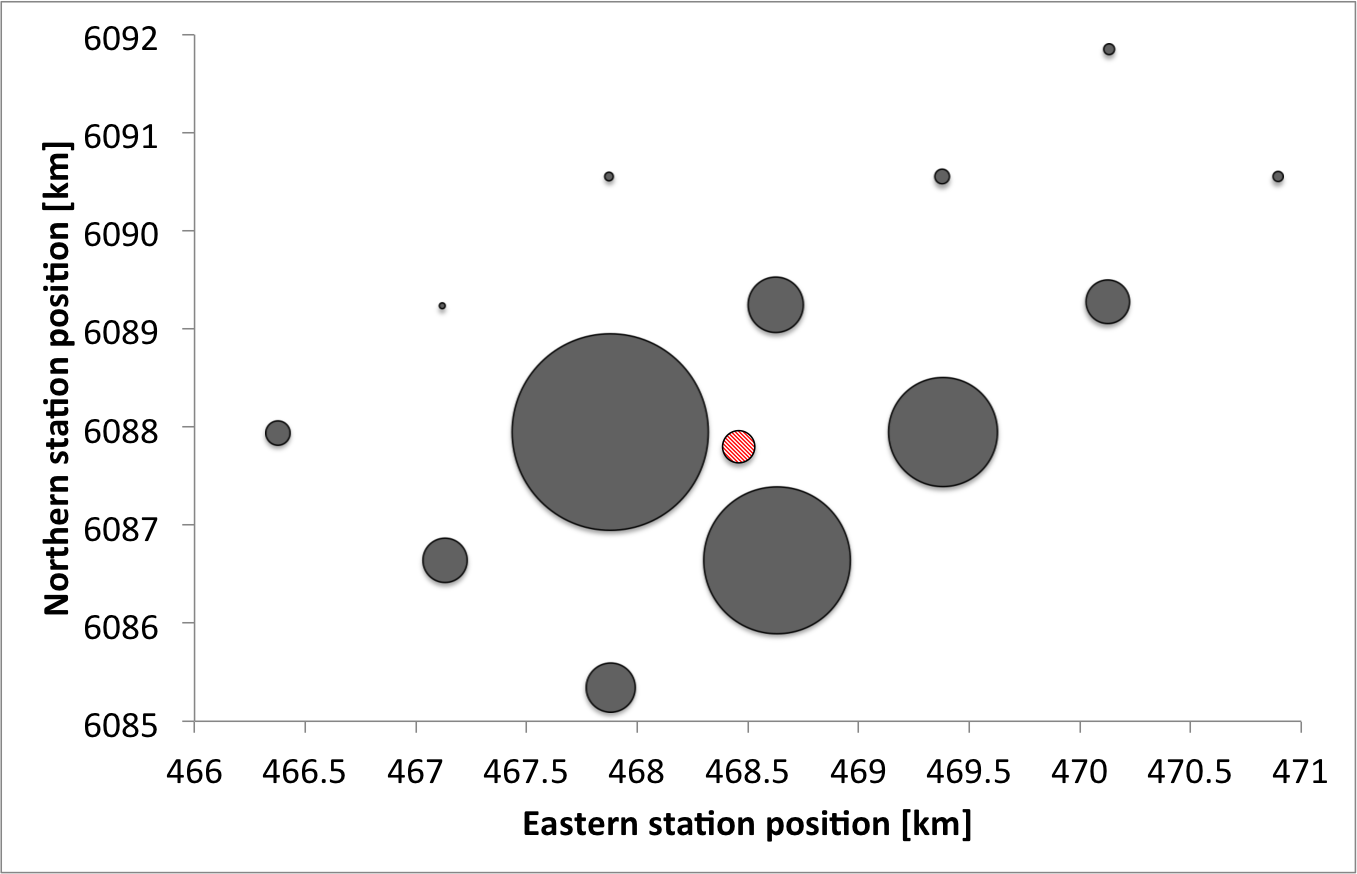}}
 \quad
 \subfigure{\includegraphics[width=0.48\textwidth]{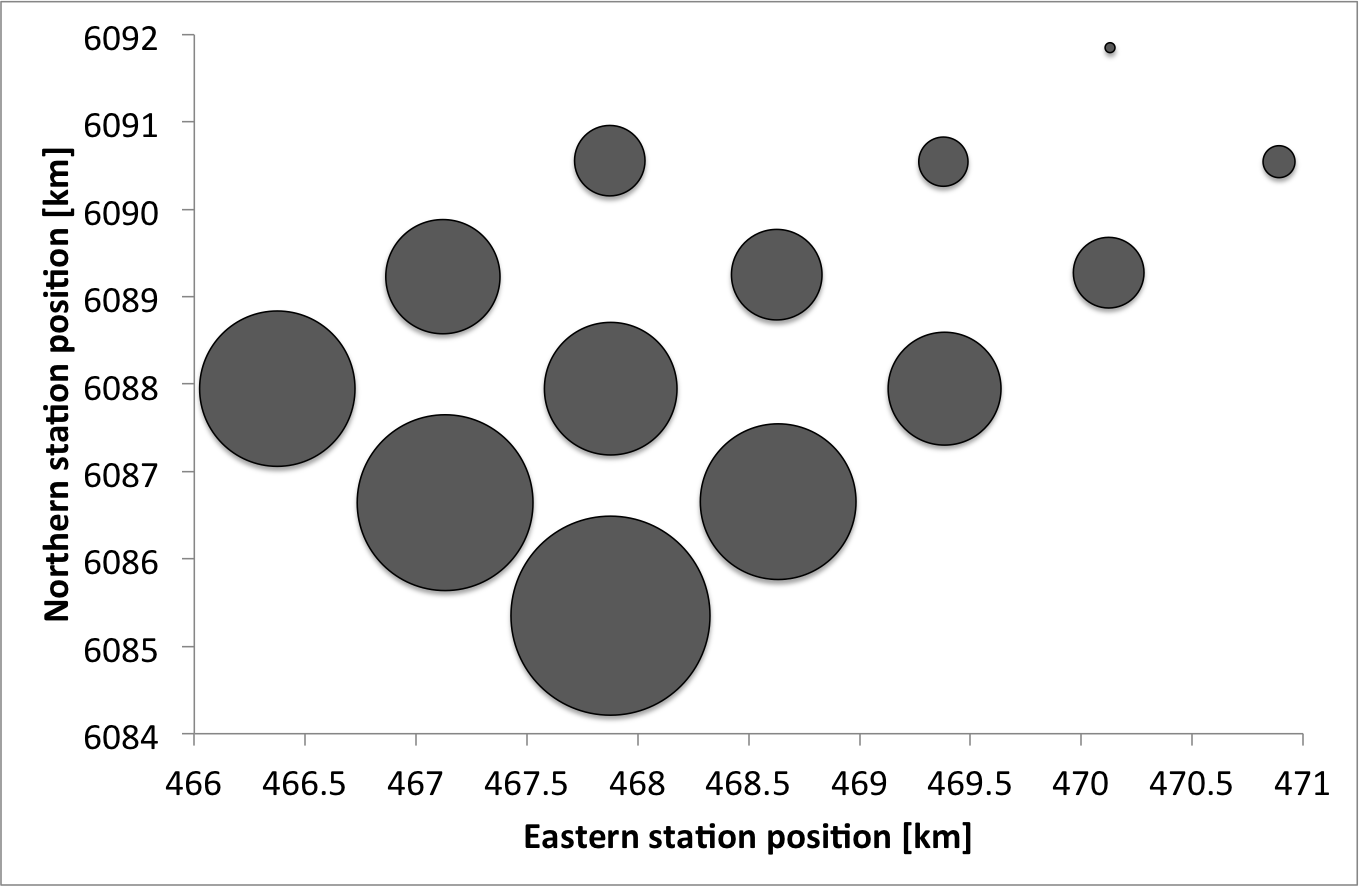}}
 \caption{\textbf{Left:} Representation of the positions of the surface detector stations hit by a shower and their corresponding signals. The center of each circle is the position of a station while the radius of the circle is proportional to the detected signal intensity in each station. The red-white shaded circle represents the position of the shower core position. \textbf{Right:} Representation of the arrival times in each station. Larger circles indicate later arrival times.}
 \label{fig:event position and time}
\end{figure}

The Pierre Auger Collaboration released today about 40000 events on public webpages \cite{PAOpublicDE, PAOpublicINT}. These events cover an energy range between \unit{0.1}{\exa eV}\footnote{\unit{1}{EeV}=\unit{10^{18}}{eV}} and \unit{49.7}{\exa eV}. Table \ref{tab:event} shows a data set of one event (ID: 03330400) which can be obtained from these web pages. The impact point can be directly reconstructed by calculating the center of mass of the detectors, with weight given by the signal using equations:

\begin{equation}
X_{\textnormal{center-of-mass}}=\frac{\sum_{i=1}^{N}X_{i}\cdot S_{i}}{\sum_{i=1}^{N}S_{i}}
\end{equation}

\begin{equation}
Y_{\textnormal{center-of-mass}}=\frac{\sum_{i=1}^{N}Y_{i}\cdot S_{i}}{\sum_{i=1}^{N}S_{i}}
\end{equation}

\noindent
where the variables $X_{i}$ and $Y_{i}$ denote the positions of the stations on the ground in UTM coordinates and $S_{i}$ is the measured signal. The result of the calculation for the example event shown in Table \ref{tab:event} is $X_{\textnormal{center-of-mass}}\approx\unit{468848}{m}$ and $Y_{\textnormal{center-of-mass}}\approx\unit{6087798}{m}$, which is marked as a shaded circle in the left-hand plot of Figure \ref{fig:event position and time}. 

\begin{table}[t]
\centerline{
\begin{tabular}{|c|c|c|c|c|}
\hline
Station ID	&		Easting position X {[}m{]}		&		Northing position Y {[}m{]}		&	Signal {[}VEM{]}		&	Time {[}ns{]}	\\
\hline
\hline
	150		&			467878.30						&		6087951.94								&		79.11				&		412891827	\\
\hline
	143		&			468632.71						&		6086649.48								&		59.13				&		412894155	\\
\hline
	144		&			469382.27						&		6087951.52								&		43.92				&		412889838	\\
\hline
	102		&			468627.23						&		6089249.33								&		22.30				&		412887588	\\
\hline
	149		&			467878.43						&		6085350.10								&		19.81				&		412898525	\\
\hline
	148		&			467132.16						&		6086643.91								&		17.85				&		412896183	\\
\hline
	105		&			470122.31						&		6089278.38								&		17.57				&		412885584	\\
\hline
	158		&			466376.12						&		6087945.37								&		9.81					&		412894124	\\
\hline
	138		&			469377.96						&		6090546.53								&		5.91					&		412883425	\\
\hline
	139		&			470131.53						&		6091848.80								&		4.37					&		412879462	\\
\hline
	140		&			470893.43						&		6090547.29								&		4.11					&		412881682	\\
\hline
	131		&			467873.99						&		6090554.71								&		3.46					&		412885564	\\
\hline
	151		&			467118.74						&		6089231.88								&		2.36					&		412889975	\\
\hline
\end{tabular}
}
\caption{Event measured by the Pierre Auger Observatory as obtained from the public webpages \cite{PAOpublicDE,PAOpublicINT}.}
\label{tab:event}
\end{table}

\begin{figure}[t]
 \centering
 \subfigure{\includegraphics[height=5.5cm,keepaspectratio]{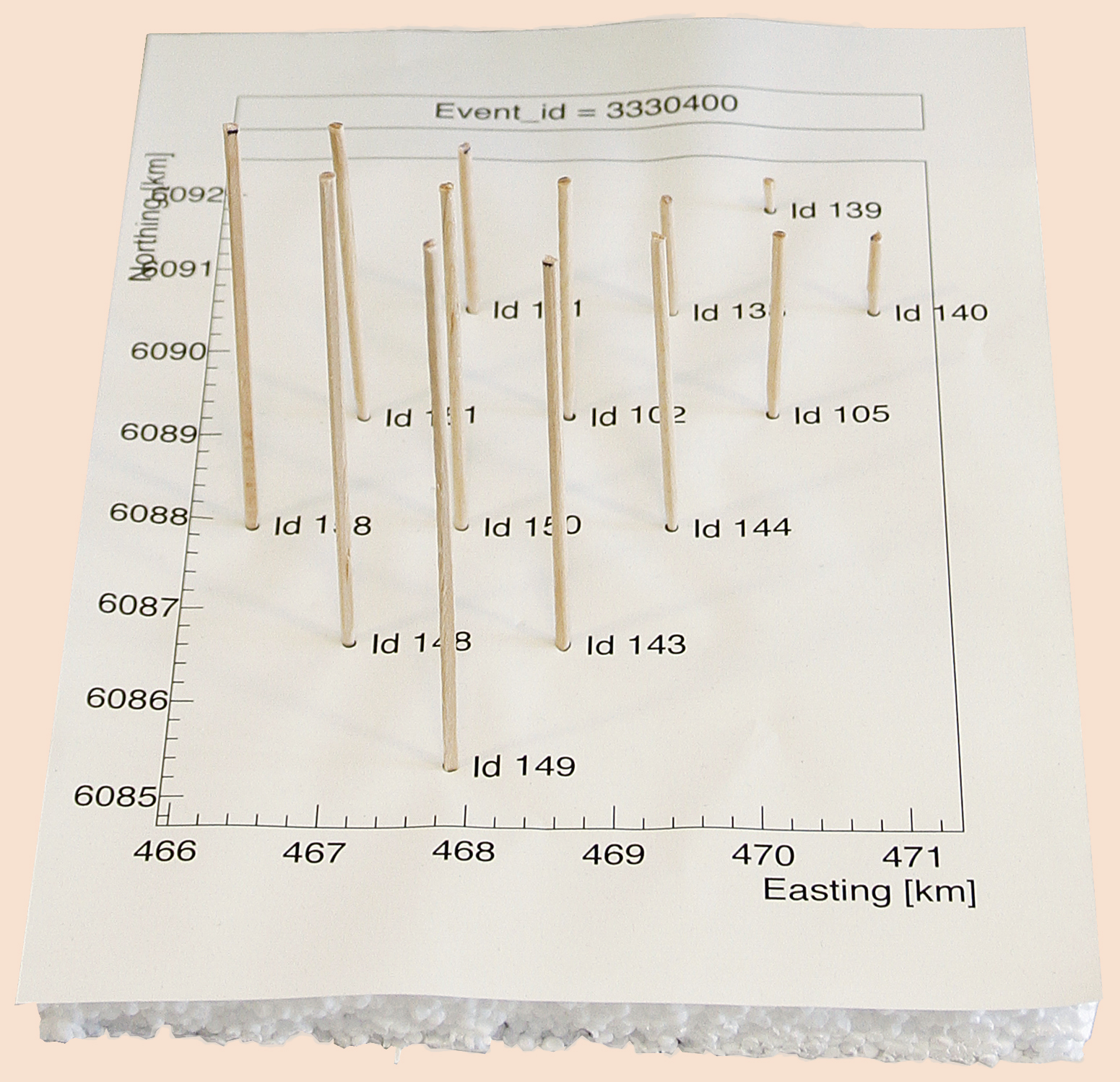}}
\quad
 \subfigure{\includegraphics[height=5.5cm,keepaspectratio]{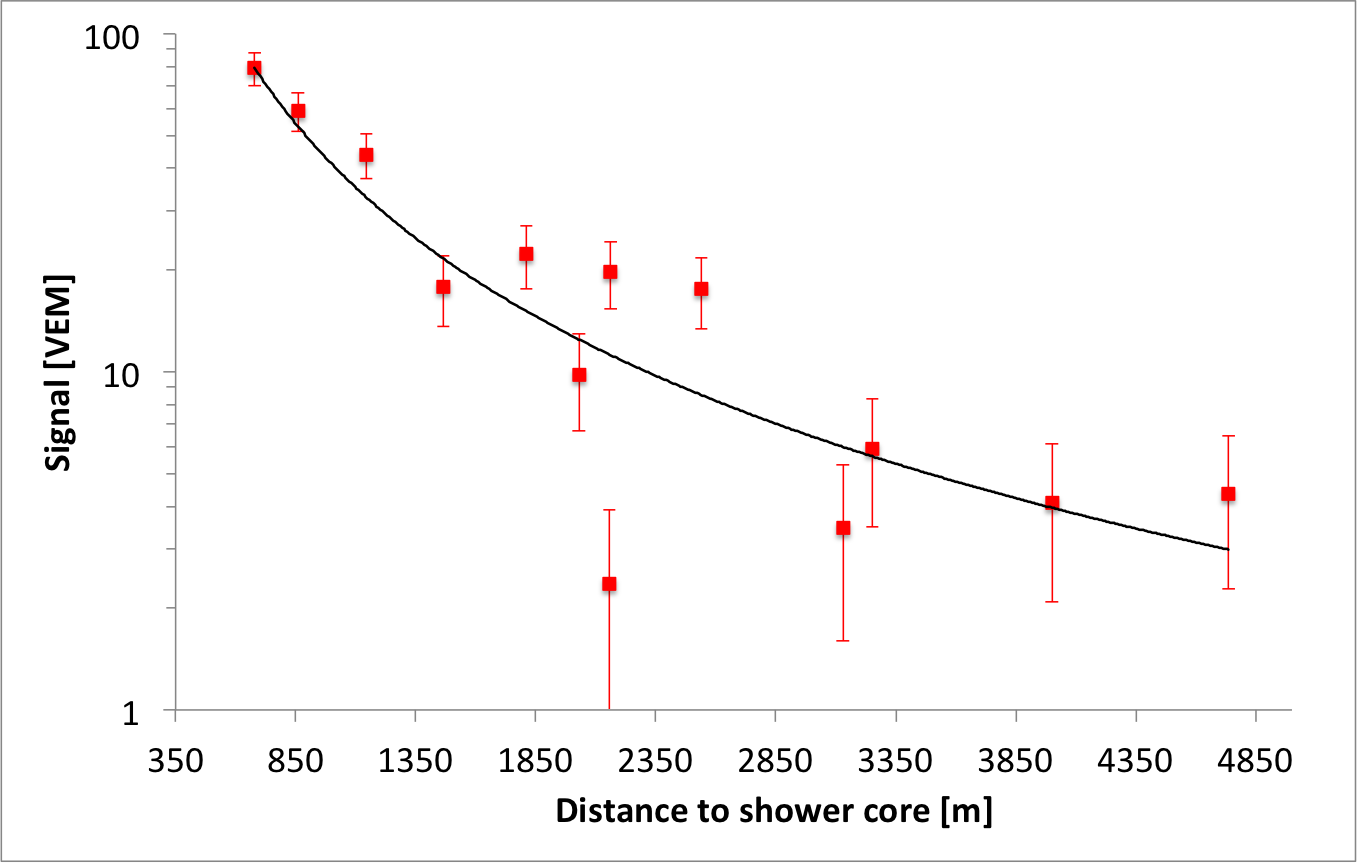}}
 \caption{\textbf{Left:} Three-dimensional model using wooden sticks to illustrate the shower front. A wooden pick marks the position of each hit station. The lengths of the sticks represent the arrival times of the EAS at each station. \textbf{Right:} The lateral shower profile of particles in an air shower with an energy of \unit{9.66}{\exa eV}.}
 \label{fig:LDF}
\end{figure}

The direction of the shower is determined by two angles, the azimuth angle $\phi$ and the zenith angle $\theta$. While the former illustrates the direction in the XY-plane, the latter indicates the inclination of the EAS. The student is guided to cut wooden dowels (e.g. supermarket shish kabob sticks) with their lengths proportional to the arrival time of the air shower in each station \cite{deSouza:2015a}. Then, a print-out of the left panel of Figure \ref{fig:event position and time} is attached to a \unit{2}{cm} styrofoam. Finally, the dowels are fixed to each station. After the sticks are fixed, it becomes apparent that they form a plane which is the shower front. This model is illustrated in Figure \ref{fig:LDF} (left). The direction of the primary particle is the direction perpendicular to the shower front. This could be easier demonstrated by laying a paper or transparency on top of the sticks. Then, the student places a stick through the impact point of the shower, which is perpendicular to the plane of the shower front. At the end, the student can measure the angle between the perpendicular stick and the ground. This is the elevation angle. In addition, the azimuth angle can be derived by measuring the angle between the projection of the stick onto the ground and the East direction. 

\begin{figure}[t]
 \centering
 \subfigure{\includegraphics[height=5cm,keepaspectratio]{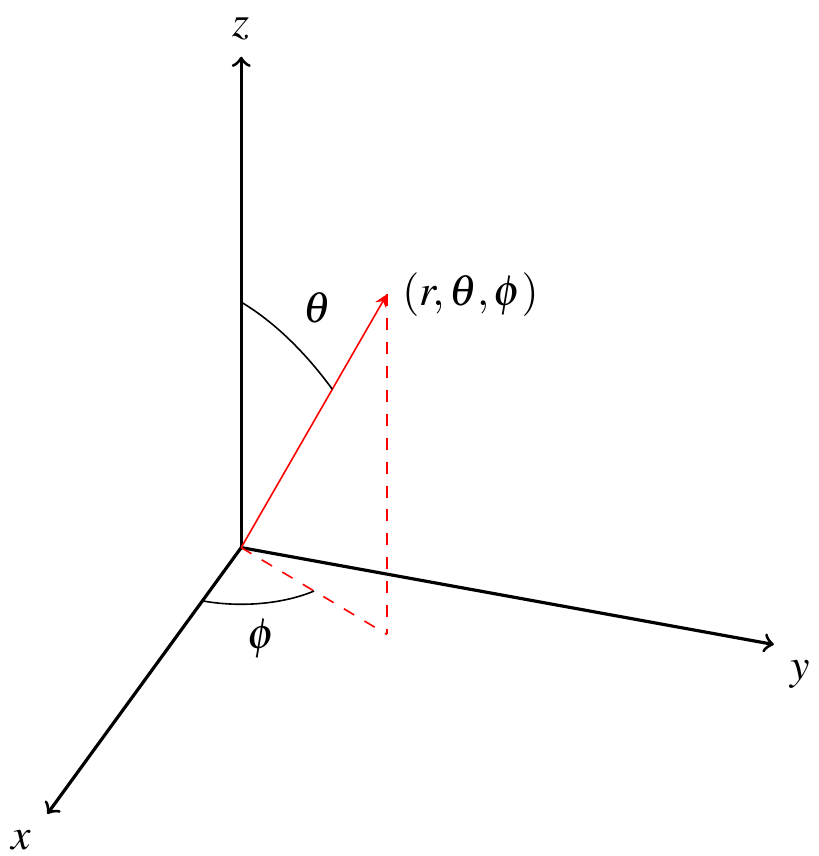}}
\quad
 \subfigure{\includegraphics[height=5cm,keepaspectratio]{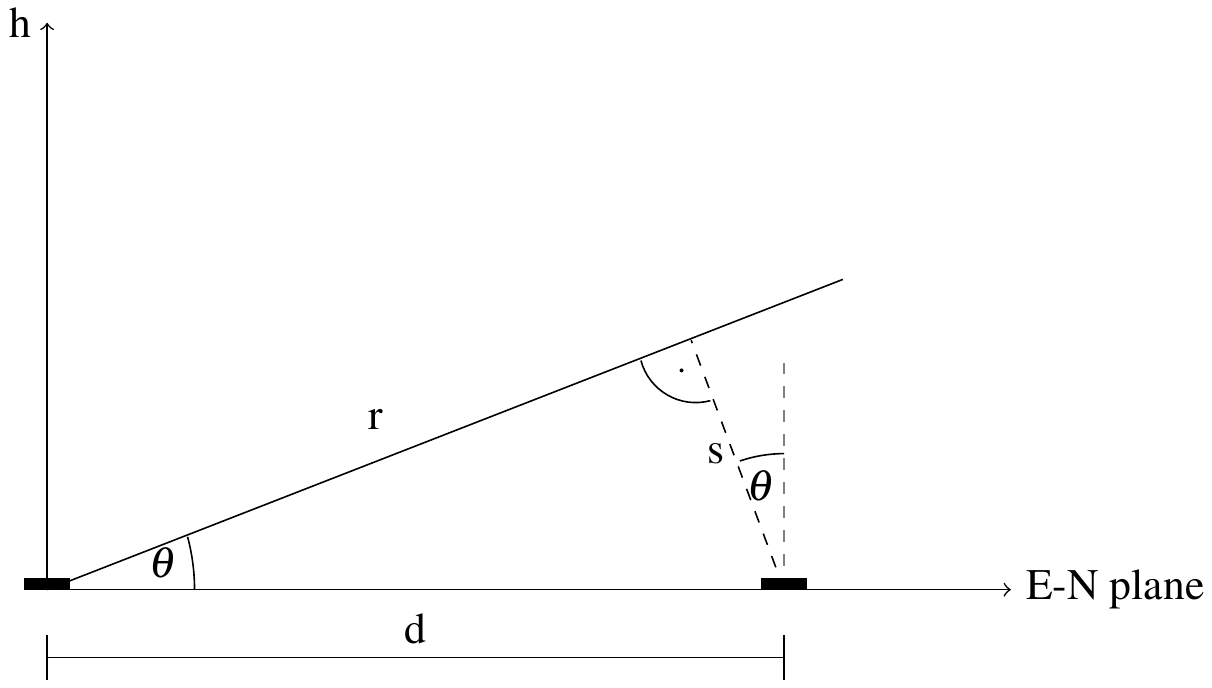}}
 \caption{The azimuth angle $\phi$ and zenith angle $\theta$ of the direction of the incoming air shower can be derived using these coordinate systems and the time of arrival of the signal in each water-Cherenkov detector station.}
 \label{fig:geometry}
\end{figure}

In addition, both angles can be calculated via a geometrical projection as illustrated in \mbox{Figure \ref{fig:geometry}}. Together with the right-hand plot of Figure \ref{fig:event position and time} the direction of the shower in the XY-plane can be estimated. This exercise is an important pedagogical tool to study mathematics in three dimensions. All stations located on a line perpendicular to the shower axis should receive a signal at the same time. The student can compute a plane equation along the three-dimensional points. Then, a linear equation which is perpendicular to this plane can be calculated. This line equation should cross the impact point in the XY-plane. This can be projected into the XY-plane which allows the student to compute the azimuth angle as the angle between the projection and the X-axis pointing East as illustrated in Figure \ref{fig:geometry}. The zenith angle will be reconstructed in the following way. The plane representing the shower front will be simplified to a plane propagating with the speed of light $c$. The incoming plane hits the first detector at a time $t_0$. To reach the next station the shower front has to travel the distance $s$, whose value will be computed from the difference in time between the two stations:
\begin{equation}
s=c\cdot\Delta t
\end{equation}

Thus, the zenith angle will be evaluated with
\begin{align} 
  \sin \theta    & =\frac{c\cdot\Delta t}{d} \nonumber \\ 
  \theta          & = \arcsin\left(\frac{c\cdot\Delta t}{d}\right)
\end{align}

Finally, the lateral distribution of the air shower will be investigated. Figure \ref{fig:LDF} (right) shows the density of the charged particles inside the shower, as a function of distance along the shower core. Using all public events with very high energies above \unit{3}{\exa eV}, the students can investigate the directions these cosmic rays come from and their distributions. They can address fundamental questions, such as the isotropy of cosmic rays or the origin of the cosmic rays measured at the Pierre Auger Observatory. In addition, students are able to establish the relationship between the number of water-Cherenkov detector stations with signals and the shower energy.

\section{Discussion}
\label{sec:discussion}

The \textit{Astroparticle Masterclass} was first organized in 2012. Since then, the number of participating high school students has increased. Table \ref{tab:masterclasses statistics} lists details on the number of events and students that participated to date. 

\begin{table}[h!]
\centerline{
\begin{tabular}{|c|c|c|}
\hline
Year 								&		Number of events		&		Students	\\
\hline
\hline
	2012							&			1								&		12	\\
\hline
	2013							&			1								&		60	\\
\hline
	2014							&			4								&		91	\\
\hline
	2015 (until 06/2015)	&			4								&		106	\\
\hline
\end{tabular}
}
\caption{\textit{Astroparticle Masterclasses} in Germany since its development in 2012.}
\label{tab:masterclasses statistics}
\end{table}

The topics that have been taught during the masterclass included the discovery of cosmic rays, the standard model of particle physics, the physics of extensive air showers, and astroparticle detectors. After the end of the activity, the students were asked to answer a short questionnaire in order to evaluate some aspects of the activity. The focus was not on the systematic study of the results. Only a first feedback from the students should be obtained to evaluate the impact of the activity and to provide information on topics to be included, improved and simplified. Overall, the results of the surveys concluded that the \textit{Astroparticle Masterclass} has been a great success. The experimental work with the data was especially met with a positive response. The requests for activities like this is increasing. Both students and teachers show a rising interest in scientific research and expanding their views by using interdisciplinary approaches. In addition, they get to know research fields which are not part of the standard curriculum at high schools. Nevertheless, the surveys show that there is still room for improvement. Sometimes, the amount of theory which is introduced in the beginning of the activity is found rather difficult to grasp. At the same time, practical experiments and their understanding are preferred by the students. In addition, students would like to know how to become a scientist after finishing high school, and to get an insight into the daily work of a researcher. 

\section{Final Remarks}
This proceeding presents an activity on astroparticle physics developed for high school students, from ages 15 to 19. It can be conducted at both schools and scientific institutes. The experiment described here allows students to reconstruct basic properties of air showers, such as the arrival direction, using simple concepts of classical physics alongside high school mathematics. Multiple aspects from a variety of subjects can be discussed and trained. These include the center-of-mass problem, as well as the usage of a spreadsheet program such as Microsoft Excel or LibreOffice Calc. 
The received feedback from the students and teachers was very positive. They show that this activity has the potential to motivate the participants in current research and physics. Furthermore, students and teachers have the chance to discuss the proposed subject in a different framework than a typical lesson at school. New activities are currently being developed to make the project less monotonous. These include an astroparticle quiz and crossword puzzle based on the topics learned during the day. Prizes will then be awarded to the winner and runner-up.

\vspace{0.7cm}
The authors acknowledge the support from \textit{Netzwerk Teilchenwelt}, which is managed by the Technical University Dresden under the umbrella of the German Physical Society (DPG). We also gratefully acknowledge the financial support of the German Ministry for Education and Research (BMBF).

\bibliographystyle{JHEP}
\bibliography{AstroParticleMasterclass}

\end{document}